\documentclass[a4paper]{article}
\usepackage[14pt]{extsizes} 
\usepackage[utf8]{inputenc}
\usepackage{listings}
\usepackage{hyperref}
\usepackage{tikz-cd}
\usepackage{mathrsfs}
\usepackage{setspace,amsmath}
\usepackage{amssymb}
\usepackage{tabularx}
\usepackage{breqn}
\usepackage{graphicx}
\usepackage{float}
\usepackage{braket}
\usepackage{comment}
\usepackage{wrapfig}
\usepackage[left=20mm, top=15mm, right=15mm, bottom=15mm, nohead, footskip=10mm]{geometry}
\usepackage{indentfirst}
\usepackage{amsmath}
\usepackage{verbatim}
\usepackage{longtable}
\usepackage{caption}
\usepackage{authblk}

\begin{document}

	\title{Construction of Mirror Pairs of Berglund-Hübsch Calabi-Yau Orbifolds of Loop-Type Polynomials}

	\author[1,2]{Maxim~Malyutin\thanks{E-mail: msmalyutin@gmail.com}}

	\affil[1]{Moscow Institute of Physics and Technology, Dolgoprudnyi, 141700, Russia}
	\affil[2]{Landau Institute for Theoretical Physics, Chernogolovka, 142432, Russia}


	\maketitle

	\begin{abstract}
		In this paper, an algorithm for constructing orbifolds of Calabi-Yau manifolds of the Berglund-Hübsch (BH) type and their mirror pairs for loop-type polynomials is proposed and applied. A complete enumeration of loop polynomials satisfying the Calabi-Yau condition was carried out, yielding 216 unique polynomials. For each polynomial, the Euler characteristics, the numbers of generations and singlets were computed, and the twisted sector states were analyzed using Roan's generalized combinatorial construction. It was established that Roan pairs are absent for all studied configurations, indicating a zero contribution of the twisted sectors to the difference in Hodge numbers upon resolving the orbifold singularities. The primary physical result of this work is the discovery of two configurations with diagonal exponents $\{74, 3, 3, 5, 3\}$ and $\{74, 3, 5, 3, 3\}$, for which the number of quark and lepton generations is precisely three ($N_{\text{gen}} = 3$), which is of key interest for constructing phenomenologically realistic models of particle physics.
	\end{abstract}

	\section{Introduction}
	
	One of the fundamental approaches to constructing realistic particle physics models within superstring theory is the compactification of the 10-dimensional heterotic string \cite{Gross1984} on 6-dimensional Calabi-Yau manifolds \cite{Candelas1985}. Such a procedure leads to 4-dimensional effective theories with $\mathcal{N}=1$ spacetime supersymmetry and an $E_6 \times E_8$ gauge group. In order for the low-energy theory to reproduce the observed properties of the physical world--in particular, precisely three generations of quarks and leptons--it is necessary to thoroughly investigate the topological and geometric properties of such spaces.
	
	In a recent paper by Belavin~A.~A. \cite{BELAVIN2025117055}, the construction of Gepner conformal models was generalized to arbitrary BH type Calabi–Yau manifolds using the Batyrev-Borisov combinatorial approach \cite{batyrev1993dualpolyhedramirrorsymmetry, borisov2011berglundhubschmirrorsymmetryvertex}. In this framework, the physical spectrum of the theory (including the numbers of gauge multiplets in the $27, \overline{27}$ representations and $E_6$ singlets) is directly determined by the combinatorial data of Batyrev reflexive polyhedra.
	
	In \cite{Aleshin2026, aleshin2026CY}, an algorithm for constructing all type orbifolds was proposed. In particular, Fermat-type examines in datail. In \cite{Aleshin:2026lep}, this approach was implemented for the case of Fermat-type diagonal potentials. These studies yielded a complete classification of Fermat-type polynomial pairs satisfying the Calabi-Yau condition, constructed their BH orbifolds, and computed their stringy Euler characteristics. Furthermore, a combinatorial procedure for computing Roan pairs \cite{PairsofRoan} and Hodge-Roan numbers was developed, taking into account the contributions of twisted sectors during singularity resolution, and its exact agreement with the results calculated via Vafa's formula \cite{Vafa:1989xc} was proved.
	
	The goal of the present work is to extend this algorithm to the class of non-diagonal potentials--loop-type polynomials. In this paper, we perform a complete enumeration of loop polynomials satisfying the Calabi-Yau condition, analyze the structure of admissible diagonal symmetry groups, construct BH mirror orbifolds, and compute their Euler characteristics. To describe the twisted sector contributions arising during the resolution of orbifold singularities, we generalize Roan's combinatorial construction from the diagonal Fermat matrix case to the general exponent matrix of a loop potential and verify the resulting recipe by comparing it with Vafa's formula.
	
	\section{The Berglund-Hübsch mirror \\ construction}
	
	We define Calabi-Yau manifolds as hypersurfaces in weighted projective space with weights $\mathbf{k}=(k_1,k_2,k_3,k_4,k_5)$ \cite{Berglund_1993}
	\begin{equation}
		\mathbb{P}_{\mathbf{k}} = \biggl\{ (x_1,\dots,x_5) \in \mathbb{C}^5 \setminus \{0\} \Big| x_i \sim \lambda^{k_i} x_i, \; \forall \lambda \in \mathbb{C}^* \biggr\}.
	\end{equation} 
	
	The hypersurface is defined by the vanishing of the polynomial
	\begin{equation}
		\label{eq:W0}
		W_0 = \sum_{i=1}^5 \prod_{j=1}^5 x_j^{A_{ij}} = 0.
	\end{equation}
	
	Here, the exponent matrix $A_{ij}$ is subject to conditions ensuring that the zero locus of polynomial \eqref{eq:W0} defines a Calabi-Yau manifold:
	
	1. The matrix $A_{ij}$ is integer-valued, non-negative, and invertible.
	
	2. The polynomial $W_0$ is quasihomogeneous, i.e., $W_0(\lambda^{k_i} x_i) = \lambda^d W_0(x)$, which, combined with the Calabi–Yau condition $d = \sum_j k_j$, yields
	\begin{equation}
		\label{eq:CY}
		\sum_{j=1}^5 A_{ij} q_j = \sum_{j=1}^5 q_j = 1, \quad q_i = \frac{k_i}{d}.
	\end{equation}
	
	3. The polynomial $W_0$ is non-degenerate away from the origin. The types of such polynomials are:
	\begin{itemize}
		\item Fermat: $x_1^{a_1} + \dots + x_n^{a_n}$
		\item Chain: $x_1^{a_1}x_2 + \dots + x_n^{a_n}$
		\item Loop: $x_1^{a_1}x_2 + \dots + x_n^{a_n}x_1$
	\end{itemize}
	as well as types formed by mixing these three. In total, there are 16 polynomial types described in \cite{Kreuzer1992}.
	
	The full family of Calabi-Yau manifolds is obtained by deforming the initial polynomial $W_0$ with monomials of the form $\prod_{i=1}^5 x_i^{S_{ji}}$. The family of Calabi-Yau manifolds is then given by
	\begin{equation}
		\mathcal{Q} = \biggl\{ (x_1,\dots,x_5) \in \mathbb{P}_{\mathbf{k}} \Big| W_0 + \sum_{j=1}^{h_X} \phi_j \prod_{i=1}^5 x_i^{S_{ji}} = 0 \biggr\},
	\end{equation}
	where the parameters $\phi_i$ are the complex structure moduli of the manifold $\mathcal{Q}$, $h_X$ is the Hodge number of the family $\mathcal{Q}$, and the matrix $S_{ji}$ satisfies the condition \eqref{eq:CY}:
	\begin{equation}
		\label{eq:def}
		\sum_{i=1}^5 S_{ji} q_i = \sum_{i=1}^5 q_i = 1,
	\end{equation}
	along with an additional condition imposed on loop-type polynomials:
	\begin{equation}
		0 \le S_{ii} \le A_{ii} - 1.
	\end{equation}
	
	To find orbifolds and construct mirror pairs, it is necessary to determine the symmetry groups of the polynomial.
	
	We define the maximal group of diagonal automorphisms preserving the original polynomial $W_0$ as
	\begin{equation}
		\text{Aut}(A) = \left\{ (e^{2\pi i g_1}, \dots, e^{2\pi i g_5}) \in (\mathbb{C}^*)^5 \Big| \sum_{j=1}^5 A_{ij} g_j \in \mathbb{Z} \right\},
	\end{equation}
	where the elements satisfy
	\begin{equation}
		g_j = \sum_{i=1}^5 B_{ji} m_i, \quad m_i \in \mathbb{Z}, \quad B = A^{-1}.
	\end{equation}
	
	The generators of the group $\text{Aut}(A)$ take the form
	\begin{equation}
		\left( e^{2\pi i B_{1j}}, \dots, e^{2\pi i B_{5j}} \right), \quad j = 1, \dots, 5.
	\end{equation}
	
	Consider the subgroup of $\text{Aut}(A)$ that preserves the monomial $\prod_{i=1}^5 x_i$:
	\begin{equation}
		SL_W = \left\{ (e^{2\pi i g_1}, \dots, e^{2\pi i g_5}) \in (\mathbb{C}^*)^5 \Big| \sum_{j=1}^5 A_{ij} g_j \in \mathbb{Z}, \; \sum_{i=1}^5 g_i \in \mathbb{Z} \right\},
	\end{equation}
	where an additional condition is imposed on the elements:
	\begin{equation}
		\sum_{j,i=1}^5 B_{ji} m_i \in \mathbb{Z}, \quad 0 \le m_j < \max_{i \in \{1,\dots,5\}} \left( \frac{1}{(B^{-1})_{ji}} \right).
	\end{equation}
	
	Within the maximal admissible group, one can single out the minimal (quantum) subgroup:
	\begin{equation}
		J_A = \left\langle \left( e^{2\pi i q_1}, \dots, e^{2\pi i q_5} \right) \right\rangle.
	\end{equation}
	
	The goal is to construct orbifolds relative to which mirror pairs will be determined. The orbifold $X$ is defined as the quotient of the Calabi-Yau manifold $\mathcal{Q}$ by a certain phase symmetry subgroup $\widetilde{G}$, i.e., $X(A,G) = \mathcal{Q} / \widetilde{G}$, where $\widetilde{G} = G / J_A$ and
	\begin{equation*}
		G = \left\{ \rho_s \in SL_W \; \Bigg| \; \rho_s \prod_{j=1}^5 x_j^{S_{ij}} = \prod_{j=1}^5 x_j^{S_{ij}} \right\}.
	\end{equation*}
	
	The group $G$ is a subgroup of $SL_W$ that preserves the admissible deformations of a given family $\mathcal{Q}$, i.e., $J_A \subset G \subset SL_W$.
	
	According to the BH construction, mirror pairs to the original Calabi-Yau manifolds are constructed using the transposed matrix $A^T$, which defines a polynomial in another weighted projective space with weights $\bar{\mathbf{k}}$:
	\begin{equation}
		\label{eq:defT}
		\sum_{j=1}^5 A_{ji} \bar{q}_j = \sum_{j=1}^5 \frac{\bar{k}_j}{\bar{d}} = \sum_{j=1}^5 \bar{q}_j = 1.
	\end{equation}
	
	The polynomial with complex structure deformations takes the form
	\begin{equation}
		W_{A^T} = \sum_{i=1}^5 \prod_{j=1}^5 x_j^{A^T_{ij}} + \sum_{i=1}^{h_Y} \psi_i \prod_{j=1}^5 x_j^{L_{ij}}.
	\end{equation}
	
	For the transposed polynomial $W_{A^T}$ one can define all the previously introduced groups. The mirror group $G^T$ is the BH dual of $G$ \cite{krawitz2009}:
	\begin{equation}
		G^T=\Bigl\{\bar\rho(\vec a)=\bigl(e^{2\pi i\sum_{j}a_jB_{ji}}\bigr)_{i=1}^5
		\;\Big|\; x^{\vec a}\ \text{is $G$-invariant}\Bigr\},
		\qquad J_{A^T}\subset G^T\subset (SL_W)^T .
	\end{equation}
	The family of deformations of $W_{A^T}$ is taken in full, and the deformations that survive in the orbifold are those invariant under $G^T$. The monomial $\prod_{j}\bar x_j^{L_{mj}}$ is $G^T$-invariant if 
	\begin{equation}
		(\vec{S}_l, \vec{L}_m) = \sum_{i,j=1}^5 S_{li} B_{ij} L_{mj} \in \mathbb{Z}
		\quad \text{for all $\vec S_l$ such that $x^{\vec S_l}$ is $G$-invariant.}
	\end{equation}
	This is the duality (pairing) condition between the $G$-invariant deformations of $W_A$ and the $G^T$-invariant deformations of $W_{A^T}$. Here $L_{mj}$ runs over the admissible deformations of $W_{A^T}$, i.e. non-negative integer	exponents satisfying the quasihomogeneity condition \eqref{eq:def}, of which there are finitely many.
	
	The orbifold $X(A^T, G^T) = \mathcal{Q}_{A^T} / \widetilde{G}(A^T)$, where $\widetilde{G}(A^T) = G(A^T) / J_{A^T}$, forms a mirror pair with $X(A, G) = \mathcal{Q} / \widetilde{G}(A)$.
	
	\section{Complete classification of loop polynomials}
	\label{sec:Class}
	A loop polynomial corresponds to an exponent matrix of the form
	\begin{equation}
		\label{eq:matrixA}
		A = \begin{pmatrix}
			a_1 & 1 & 0 & 0 & 0 \\
			0 & a_2 & 1 & 0 & 0 \\
			0 & 0 & a_3 & 1 & 0 \\
			0 & 0 & 0 & a_4 & 1 \\
			1 & 0 & 0 & 0 & a_5
		\end{pmatrix},
		\qquad a_i \in \mathbb{Z}_{\ge 2}.
	\end{equation}
	
	Condition \eqref{eq:CY} can be written as
	\begin{equation}
		\label{eq:RelLoop}
		a_i q_i + q_{i+1} = 1,
		\qquad i = 1, \dots, 5,
		\qquad q_6 = q_1.
	\end{equation}
	The Calabi-Yau condition \eqref{eq:CY} takes the form
	\begin{equation}
		\label{eq:CYLoop}
		\sum_{i=1}^{5} q_i = 1.
	\end{equation}
	
	The system \eqref{eq:RelLoop}--\eqref{eq:CYLoop} is invariant under simultaneous cyclic permutations of the pairs $(a_i, q_i)$. Therefore, without loss of generality, we can fix
	\begin{equation}
		\label{eq:q1}
		q_5 = \max_{1 \le i \le 5} q_i.
	\end{equation}
	Then $q_5 \ge 1/5$. Since $q_1 > 0$, the first equation in \eqref{eq:RelLoop} implies
	\begin{equation}
		a_5 q_5 = 1 - q_1 < 1.
	\end{equation}
	Consequently,
	\begin{equation}
		2 \le a_5 < \frac{1}{q_5} \le 5,
	\end{equation}
	and thus
	\begin{equation}
		a_5 \in \{2, 3, 4\}.
		\label{eq:a1bounds}
	\end{equation}
	
	The Calabi-Yau condition \eqref{eq:CY} for the inverse matrix $B = A^{-1}$ is written as
	\begin{equation}
		\label{eq:eqB}
		\sum_{j=1}^{5} B_{ij} = q_i, \quad \sum_{i,j=1}^{5} B_{ij} = 1.
	\end{equation}
	
	Let us find effective bounds for the exponents $a_i$ that encompass all solutions to the Calabi-Yau condition \eqref{eq:eqB}. To this end, we introduce the substitution
	\[
	a_1 = x + 2, \qquad a_2 = t + 1, \qquad a_3 = z + 2, \qquad a_4 = w + 2,
	\]
	where $x \in \{0, 1, 2\}$, $t \ge 1$, and $z, w \ge 0$, thereby reducing equation \eqref{eq:eqB} to the form
	\begin{equation}
		a_5 = \frac{tE + F}{tP - H},
		\label{eq:a5ratio}
	\end{equation}
	where
	\begin{align*}
		P &= wxz + wx + wz + xz + x - 2, \\
		E &= wxz + 2wx + wz + 2w + xz + 3x + z + 3, \\
		F &= wxz + 2wx + 2wz + 3w + xz + 2x + 2z + 5, \\
		H &= w + xz + 2x + 2z + 5.
	\end{align*}
	
	For admissible values of the exponents, $P > 0$, and the right-hand side of \eqref{eq:a5ratio} decreases monotonically with respect to $t$. From the condition $tP - H > 0$, it follows that the minimum admissible integer value of $t$ is
	\begin{equation}
		t_{\text{min}} = \left\lfloor \frac{H}{P} \right\rfloor + 1 = n + 1.
	\end{equation}
	
	Direct verification using exact integer arithmetic in Mathematica shows that $H/P \le 13$ for the three values of $x$:
	\begin{equation}
		13P - H = \begin{cases}
			13wz - w - 2z - 31, & x = 0, \\
			2(13wz + 6w + 5z - 10), & x = 1, \\
			39wz + 25w + 22z - 9, & x = 2.
		\end{cases}
	\end{equation}
	Given $P > 0$ and $z, w \ge 0$, all three expressions are non-negative; therefore, it suffices to consider $n = 0, \dots, 13$. Since the right-hand side of \eqref{eq:a5ratio} is monotonically decreasing with respect to $t$, its maximum value is attained at $t = n + 1$. Consequently, it is sufficient to check the inequality
	\begin{equation}
		\frac{(n+1)E + F}{(n+1)P - H} \le 165.
	\end{equation}
	
	Exact integer verification using exact computer arithmetic for $n = 0, \dots, 13$ demonstrates that this inequality holds. Consequently, $a_5 \le 165$. By the cyclic symmetry of $a_i$, it follows that $165$ serves as an upper bound for any element $a_i$.
	
	Thus, for a complete enumeration, it is sufficient to consider integer tuples of exponents satisfying
	\begin{equation}
		2 \le a_5 \le 4,
		\qquad
		2 \le a_i \le 165,
		\qquad i = 1, 2, 3, 4.
		\label{eq:LoopBounds}
	\end{equation}
	For each such tuple, condition \eqref{eq:eqB} is verified.
	
	Two exponent tuples are identified if they are related by a cyclic renumbering of the variables:
	\begin{equation}
		(a_1, a_2, a_3, a_4, a_5)
		\sim
		(a_2, a_3, a_4, a_5, a_1).
		\label{eq:C5equiv}
	\end{equation}
	
	After identifying cyclically equivalent tuples, we obtain 216 classes of loop polynomial exponents satisfying the Calabi-Yau condition. The resulting classes are presented in Section~\ref{sec:Result}.
	
	\section{Structure of symmetry groups}
	
	Let us write the elements of the diagonal automorphism group in terms of their exponent phase shifts:
	
	\begin{equation}
		\text{Aut}_{\mathbb{Q}/\mathbb{Z}}(A) = \left\{ g \in (\mathbb{Q}/\mathbb{Z})^5 \Big| Ag \in \mathbb{Z}^5 \right\},
	\end{equation}
	which can be rewritten as $\text{Aut}_{\mathbb{Q}/\mathbb{Z}}(A) = A^{-1}\mathbb{Z}^5 / \mathbb{Z}^5 \cong \mathbb{Z}^5 / A\mathbb{Z}^5$. Let the Smith normal form of the matrix $A$ be
	
	\begin{equation}
		UAV = D = \text{diag}(d_1, \dots, d_5),
	\end{equation}
	where $U, V$ are unimodular integer matrices and $d_1 \mid d_2 \mid \dots \mid d_5$. Hence,
	
	\begin{equation}
		\mathbb{Z}^5 / A\mathbb{Z}^5 \cong \mathbb{Z}^5 / D\mathbb{Z}^5 \cong \mathbb{Z}/d_1\mathbb{Z} \times \dots \times \mathbb{Z}/d_5\mathbb{Z}.
	\end{equation}
	
	Consider the 4th-order minor of the matrix \eqref{eq:matrixA}:
	
	\begin{equation}
		\begin{pmatrix}
			1 & 0 & 0 & 0 \\    
			a_2 & 1 & 0 & 0 \\
			0 & a_3 & 1 & 0 \\
			0 & 0 & a_4 & 1
		\end{pmatrix};
	\end{equation}
	its determinant is equal to 1, which implies that the greatest common divisor (GCD) of all 4th-order minors is 1. For the Smith normal form \cite{aleshin2026CY}, the relation $d_k = \Delta_k / \Delta_{k-1}$ holds, where $\Delta_k$ is the GCD of all $k$-th order minors, with $\Delta_0 = 1$ and $\Delta_5 = \det A$. Thus, $d_1 d_2 d_3 d_4 = \Delta_4 = 1$, and we obtain the following Smith normal form:
	
	\begin{equation}
		D = \text{diag}(1, 1, 1, 1, \det A).
	\end{equation}
	
	The maximal diagonal automorphism group
	
	\begin{equation}
		\text{Aut}_{\mathbb{Q}/\mathbb{Z}}(A) \cong \mathbb{Z} / (\det A)\mathbb{Z}
	\end{equation}
	is immediately shown to be cyclic. Since $SL_W \subset \text{Aut}_{\mathbb{Q}/\mathbb{Z}}(A)$, the maximal admissible group is also cyclic. Any quotient of a cyclic group, including $SL_W / J_A$, is likewise cyclic.
	
	\section{Euler characteristics and contributions of \\ twisted sectors}
	
	The original orbifold $\mathcal{Q}/G$ is generally a singular manifold. To calculate physical characteristics, such as the number of generations, one must pass to a smooth manifold $\widetilde{\mathcal{Q}/G}$ obtained by resolving the singularities of the original orbifold.
	
	A generalization of Vafa's formula for computing the Euler characteristic of the resolved orbifold $\widetilde{\mathcal{Q}/G}$ for loop polynomials is given by the following expression:
	
	\begin{equation}
		\label{eq:Vafa}
		\chi\left(\widetilde{\mathcal{Q}/G}\right) = \frac{1}{|G|} \sum_{g_1,g_2 \in G} \left( \prod_{i \in I(g_1,g_2)} \left(1 - \frac{1}{q_i}\right) \right),
	\end{equation}
	where $J_M \subset G \subset SL_W$, and $I(g_1,g_2)$ is the subset of indices $i \in \{1,\dots,5\}$ for which the $i$-th coordinate of both elements $g_1$ and $g_2$ is simultaneously equal to 1. Half of the absolute value of this Euler characteristic determines the number of generations $N_{\text{gen}}$:
	
	\begin{equation}
		N_{\text{gen}} = \frac{\left|\chi\left(\widetilde{\mathcal{Q}/G}\right)\right|}{2} = h_X - h_Y.
	\end{equation}
	
	When constructing BH mirror pairs, the difference between the numbers of invariant complex structure deformations (corresponding to the untwisted sector) for a manifold and its mirror partner does not always coincide with half of the Euler characteristic computed via Vafa's formula. The missing contributions to the Hodge numbers are introduced by the singularity resolution procedure. These contributions constitute the so-called twisted sector.
	
	To account for the states in the twisted sector, we employ a generalization of Roan's combinatorial construction. For loop-type polynomials, the contributions from the resolution of singularities are in one-to-one correspondence with Roan pairs.
	
	We define a Roan pair $(G^T,G)$ as a pair of elements $(\mathbf{z}',\mathbf{z}) \in G^T_{2Z} \times G_{3Z}$ that satisfies the "orthogonality" condition with respect to the exponent matrix of the loop polynomial:
	
	\begin{equation}
		(\mathbf{z}')^T A \mathbf{z} = 0,
	\end{equation}
	where $G^T_{2Z} \subset G^T$ is the subset of age-1 elements in $G^T$ having two zero coordinates, and $G_{3Z} \subset G$ is the subset of age-1 elements in $G$ having three zero coordinates.
	
	Symmetrically, a Roan pair $(G,G^T)$ is defined as pairs $(\mathbf{z},\mathbf{z}') \in G_{2Z} \times G^T_{3Z}$ satisfying the condition
	
	\begin{equation}
		\mathbf{z}^T A \mathbf{z}' = 0.
	\end{equation}
	
	The number of such pairs precisely compensates for the difference between the Hodge numbers of the untwisted sector and the full Hodge numbers of the resolved manifold, ensuring the validity of the generalized Vafa formula.
	
	\section{Algorithm for the construction of Calabi-Yau orbifolds and mirror pairs}
	\label{sec:algor}
	
	Let us define the steps of the algorithm for constructing BH Calabi-Yau orbifolds and their mirror pairs.
	
	In the first step, we determine all elements of the set of admissible deformations $\mathcal{S}$ for the initial polynomial $W_0$ using the Calabi-Yau condition:
	\begin{equation}
		\sum_{j=1}^5 S_{ij} k_j = \sum_{i=1}^5 k_i = d, \quad 0 \le S_{ii} \le A_{ii} - 1,
	\end{equation}
	subject to the condition for a loop-type polynomial.
	
	In the second step, we determine all elements of the maximal allowed group $SL_W$ using:
	\begin{equation}
		\sum_{i,j=1}^5 B_{ij} m_j \in \mathbb{Z}, \quad 0 < m_j \le \max_{i=1 \dots 5} \frac{1}{(B^{-1})_{ij}}.
	\end{equation}
	
	Next, for each element of the group $SL_W$, we find all deformations on which it acts invariantly:
	\begin{equation}
		R(g) = \left\{ S \in \mathcal{S} \;\Big|\; \sum_{j=1}^5 S_j g_j \in \mathbb{Z} \right\},
	\end{equation}
	where $S$ is an element of the set of admissible deformations found in first step, and $g \in SL_W$ is a group element. This yields a collection of deformation subsets $R^{(\alpha)}$.
	
	For each $R^{(\alpha)}$, we find all elements of the group $SL_W$ that act invariantly on every element of $R^{(\alpha)}$. The set of all such group elements for each deformation subset $R^{(\alpha)}$ forms a subgroup:
	\begin{equation}
		G(R^{(\alpha)}) = \left\{ g \in SL_W \;\Big|\; \sum_{j=1}^5 S_j g_j \in \mathbb{Z}, \; \forall S \in R^{(\alpha)} \right\}.
	\end{equation}
	It is the quotient group $G(R^{(\alpha)}) / J_A$ with respect to this subgroup that specifies the orbifold for the sector $R^{(\alpha)}$.

	We then find the elements of the quotient group $G(R^{(\alpha)}) / J_A$, where $J_A$ is the quantum group:
	\begin{equation}
		G(R^{(\alpha)}) / J_A = \left\{ [g] \;\Big|\; g \in G(R^{(\alpha)}) \right\},
	\end{equation}
	where
	\begin{equation}
		[g] = \left\{ g' \in G(R^{(\alpha)}) : g - g' \in J_A \right\}.
	\end{equation}
	
	Next, we determine all possible types of intersections of the sets $R^{(\alpha)}_{\text{sec}} = \bigcap_{\alpha} R^{(\alpha)}$, and for each intersection, we find all elements of the group $SL_W$ that act invariantly on every element of a fixed intersection type. These sets of elements form a subgroup of $SL_W$:
	\begin{equation}
		\label{eq:G(Ra)}
		G(R^{(\alpha)}) = \left\{ g \in SL_W \;\Big|\; \sum_{j=1}^5 S_j g_j \in \mathbb{Z}, \; \forall S \in R^{(\alpha)}_{\text{sec}} \right\},
	\end{equation}
	which defines the orbifold for the intersection after quotienting by the quantum symmetry group $J_A$.
	
	For each $R^{(\alpha)}$ and all possible intersection types $R^{(\alpha)}_{\text{sec}}$, we find subsets of mirror deformations such that pairing any element of these subsets with any element of $R^{(\alpha)}$ or the intersection $R^{(\alpha)}_{\text{sec}}$ yields an integer:
	\begin{equation}
		L^{(\alpha)} = L(R^{(\alpha)}) = \left\{ L \in \mathcal{L} \;\Big|\; \sum_{i,j=1}^5 S_m B_{mn} L_n \in \mathbb{Z}, \; \forall S \in R^{(\alpha)} \right\},
	\end{equation}
	where $\mathcal{L}$ is the complete set of admissible deformations of the mirror polynomial $W_{A^T}$. This yields collections of mirror deformations $L^{(\alpha)}$.
	
	Next, we find the elements of the maximal allowed group for the transposed matrix $A^T$. To do this, we transpose the matrix $A$, which specifies the new polynomial $W_{A^T}$, and find the new weights $\mathbf{\bar{k}}$. We then compute the elements of the maximal allowed group $(SL_W)^T$. For each $L^{(\alpha)}$, we find all elements of $(SL_W)^T$ that act invariantly on every element of $L^{(\alpha)}$. The set of all such group elements for each deformation subset $L^{(\alpha)}$ forms a subgroup of $(SL_W)^T$, which, after quotienting by the quantum symmetry group $J_{A^T}$, specifies the orbifold for $L^{(\alpha)}$:
	\begin{equation}
		G^T(L^{(\alpha)}) = \left\{ \bar{g} \in (SL_W)^T \;\Big|\; \sum_{m=1}^5 L_m \bar{g}_m \in \mathbb{Z}, \quad \forall L \in L^{(\alpha)} \right\}.
	\end{equation}
	
	The number of generations for a non-singular orbifold can be found via the deformations of the initial orbifold $R^{(\alpha)}$ and its conjugate $L^{(\alpha)}$:
	\begin{equation}
		\label{eq:deform}
		N_{\text{gen}} = \Big| |R^{(\alpha)}| - |L^{(\alpha)}| \Big|,
	\end{equation}
	where $|R^{(\alpha)}|$ and $|L^{(\alpha)}|$ denote the cardinalities (numbers of elements) of the corresponding deformation sets. We can identify $|R^{(\alpha)}|$ and $|L^{(\alpha)}|$ with the Hodge numbers $h_X(\mathcal{Q}/G) = |R^{(\alpha)}|$ and $h_Y(\mathcal{Q}/G) = |L^{(\alpha)}|$.
	
	If the orbifold is singular, the formula based on the difference of deformation cardinalities does not give the complete result. In this case, the number of generations can be computed via the Euler characteristic \eqref{eq:Vafa} for the orbifold with resolved singularities. An alternative combinatorial approach is to compute the Roan pairs $(G^T, G)$ and $(G, G^T)$, which characterize the states of the twisted sector, and add their numbers to the Hodge numbers of the untwisted sector (the cardinalities of the deformation sets $|R^{(\alpha)}|$ and $|L^{(\alpha)}|$, respectively):
	\begin{equation}
		h_X(\widetilde{\mathcal{Q}/G}) = h_X(\mathcal{Q}/G) + \big|\{(G^T, G)\}\big|,
	\end{equation}
	\begin{equation}
		h_Y(\widetilde{\mathcal{Q}/G}) = h_Y(\mathcal{Q}/G) + \big|\{(G, G^T)\}\big|.
	\end{equation}
	
	Then, the number of generations for the resolved orbifold $\widetilde{\mathcal{Q}/G}$ is given by the absolute difference of the total Hodge numbers:
	\begin{equation}
		N_{\text{gen}} = \big| h_X(\widetilde{\mathcal{Q}/G}) - h_Y(\widetilde{\mathcal{Q}/G}) \big|.
	\end{equation}
	
	Singlet states can be found as pairs of "orthogonal" deformations:
	\begin{equation}
		\label{eq:singlets}
		\sum_{n,m=1}^5 S_m B_{mn} L_n = 0, \quad \forall S_m \in R^{(\alpha)}, \quad \forall L_n \in L^{(\alpha)}.
	\end{equation}

	\section{Results}
	\label{sec:Result}
	
	This section presents the results of applying the algorithm described in Section~\ref{sec:algor} to 216 classes of loop polynomial exponents satisfying the Calabi-Yau condition (see Section~\ref{sec:Class}). The computations were split into two cases, depending on the relation between the quantum symmetry and the maximal admissible group of diagonal automorphisms.
	
	\subsection{The case $J_A=SL_W$}
	
	In the case where the quantum subgroup coincides with the maximal allowed group $J_A=SL_W$, 185 mirror pairs were obtained, the characteristics of which are presented in Table~\ref{tab1}. Polynomials for which the number of generations is $N_{\text{gen}}=3$ are of particular physical interest, as they can serve as a basis for constructing phenomenologically realistic particle physics models. In our sample, two such polynomials were found, given by the diagonal exponents $\{74, 3, 3, 5, 3\}$ and $\{74, 3, 5, 3, 3\}$.
	
	For all polynomials in this class, the number of generations calculated via the Euler characteristic \eqref{eq:Vafa} and via the difference in cardinalities of the sets of invariant deformations \eqref{eq:deform} coincide. It is interesting to note that Roan pairs are absent in all 185 cases, which indicates a zero contribution of twisted sectors to the difference in Hodge numbers upon resolution of singularities for these configurations.

			\subsection{The case $J_A \subset SL_W$}

			In the case where the quantum subgroup is a proper subgroup of the maximal allowed group $J_A \subset SL_W$, 31 mirror pairs were obtained. The results are presented in Table~$\ref{tab2}$. Unlike the previous case, polynomials with $N_{\text{gen}}=3$ generations are absent here.
			
			As in the case $J_A = SL_W$, the calculations of the number of generations using the two methods are in complete agreement, and Roan pairs do not arise. In Table~\ref{tab2}, $G_1$ and $G_2$ denote the factor groups that preserve a certain subclass of invariant deformations. Remarkably, the group $G_1$ turns out to be trivial (isomorphic to $\mathbb{Z}_1$) in all considered cases. This fact is a direct consequence of the maximal allowed group $SL_W$ always being cyclic.

	\section{Conclusion}
	
	In this work, the algorithm was extended to loop-type polynomials for constructing BH Calabi-Yau orbifolds and their mirror pairs. The proposed approach makes it possible to search for Calabi-Yau manifolds with a topology that yields three generations of quarks and leptons, which is of primary interest from a phenomenological perspective. As a result, all loop-type polynomials satisfying the Calabi-Yau condition were found, totaling 216. For each class, the Euler characteristics, Roan pairs, and the numbers of generations and singlets were calculated. In all cases, Roan pairs are absent, indicating a zero contribution from the twisted sector to the difference in Hodge numbers. Meanwhile, a generation number equal to three is realized for exactly two polynomials: $\{74, 3, 3, 5, 3\}$ and $\{74, 3, 5, 3, 3\}$.
	
	\section*{Acknowledgments}
	
	Author acknowledge A. Belavin and G. Koshevoy. The work is supported by the Russian Science Foundation grant 23-12-00333.

\bibliographystyle{unsrt} 
\bibliography{library} 	

\newpage
\section*{Appendix A}

\footnotesize
\begin{longtable}{|c|c|c|c|c||c|c|c|c|c|}	
	\caption{\normalsize Number of generations and singlets for the case $J_A = SL_W$. Here, $\text{diag}(A)$ denotes the diagonal elements of the loop polynomial matrix \eqref{eq:matrixA}. The Euler characteristic $\chi$ for each case is calculated using Vafa's formula \eqref{eq:Vafa}. Note that the number of generations $N_{\text{gen}}$, computed as the difference in the number of deformations \eqref{eq:deform}, is related to the Euler characteristic via $|\chi|=2N_{\text{gen}}$, indicating the absence of contributions from Roan pairs. The number of singlets is obtained using formula \eqref{eq:singlets}.}
	\label{tab1}\\	
	\hline

	\textbf{№} & \textbf{$\text{diag}(A)$} & $\chi$ & \textbf{$N_{\text{gen}}$} & \textbf{Singlets} & \textbf{№} & \textbf{$\text{diag}(A)$} & $\chi$ & \textbf{$N_{\text{gen}}$} & \textbf{Singlets} \\
	\hline
	\endfirsthead
	
	\multicolumn{10}{c}{} \\ 
	\hline
	\textbf{№} & \textbf{$\text{diag}(A)$} & $\chi$ & \textbf{$N_{\text{gen}}$} & \textbf{Singlets} & \textbf{№} & \textbf{$\text{diag}(A)$} & $\chi$ & \textbf{$N_{\text{gen}}$} & \textbf{Singlets}\\
	\hline
	\endhead
	
	\hline
	\multicolumn{10}{r}{} \\ 
	\endfoot
	
	\hline
	\endlastfoot
	
	1 & $\{3, 3, 14, 147, 2\}$ & 84 & 42 & 0 & 94 & $\{8, 4, 10, 4, 2\}$ & 184 & 92 & 12 \\
	2 & $\{3, 3, 15, 79, 2\}$ & 174 & 87 & 7 & 95 & $\{8, 5, 3, 4, 3\}$ & 120 & 60 & 21\\
	3 & $\{3, 3, 17, 45, 2\}$ & 246 & 123 & 9 & 96 & $\{8, 5, 10, 3, 2\}$ & 0 & 0 & 61\\
	4 & $\{3, 3, 21, 28, 2\}$ & 336 & 168 & 0 & 97 & $\{8, 6, 118, 2, 2\}$ & 24 & 12 & 2\\
	5 & $\{3, 3, 30, 19, 2\}$ & -300 & 150 & 0 & 98 & $\{8, 22, 6, 2, 2\}$ & -144 & 72 & 28\\
	6 & $\{3, 3, 47, 15, 2\}$ & -210 & 105 & 7 & 99 & $\{9, 3, 6, 31, 2\}$ & 252 & 126 & 17\\
	7 & $\{3, 3, 81, 13, 2\}$ & -138 & 69 & 9 & 100 & $\{9, 3, 12, 7, 2\}$ & -336 & 168 & 0\\
	8 & $\{3, 3, 149, 12, 2\}$ & -48 & 24 & 0 & 101 & $\{9, 5, 9, 3, 2\}$ & -210 & 105 & 10\\
	9 & $\{3, 4, 7, 47, 2\}$ & 96 & 48 & 14 & 102 & $\{9, 6, 3, 64, 2\}$ & 144 & 72 & 17\\
	10 & $\{3, 4, 11, 11, 2\}$ & -48 & 24 & 27 & 103 & $\{9, 8, 3, 12, 2\}$ & 288 & 144 & 5\\
	11 & $\{3, 5, 10, 7, 2\}$ & -180 & 90 & 6 & 104 & $\{9, 10, 3, 8, 2\}$ & -120 & 60 & 35\\
	12 & $\{3, 5, 85, 4, 2\}$ & 0 & 0 & 3 & 105 & $\{9, 13, 7, 2, 2\}$ & -84 & 42 & 39\\
	13 & $\{3, 6, 5, 24, 2\}$ & 120 & 60 & 34 & 106 & $\{9, 36, 3, 4, 2\}$ & 24 & 12 & 21\\
	14 & $\{3, 6, 21, 4, 2\}$ & -96 & 48 & 30 & 107 & $\{10, 3, 3, 9, 3\}$ & -174 & 87 & 12\\
	15 & $\{3, 7, 5, 14, 2\}$ & 156 & 78 & 21 & 108 & $\{10, 3, 9, 3, 3\}$ & 174 & 87 & 12\\
	16 & $\{3, 7, 9, 5, 2\}$ & -30 & 15 & 45 & 109 & $\{10, 3, 23, 5, 2\}$ & 36 & 18 & 42\\
	17 & $\{3, 8, 4, 97, 2\}$ & 24 & 12 & 5 & 110 & $\{10, 4, 4, 20, 2\}$ & -136 & 68 & 19\\
	18 & $\{3, 8, 51, 3, 2\}$ & -48 & 24 & 9 & 111 & $\{11, 3, 13, 6, 2\}$ & 252 & 126 & 8\\
	19 & $\{3, 9, 5, 9, 2\}$ & 210 & 105 & 10 & 112 & $\{11, 3, 157, 4, 2\}$ & -48 & 24 & 0\\
	20 & $\{3, 9, 29, 3, 2\}$ & -90 & 45 & 23 & 113 & $\{11, 6, 4, 5, 2\}$ & 120 & 60 & 25\\
	21 & $\{3, 10, 5, 8, 2\}$ & 0 & 0 & 61 & 114 & $\{11, 11, 4, 3, 2\}$ & 48 & 24 & 27\\
	22 & $\{3, 11, 4, 19, 2\}$ & -144 & 72 & 11 & 115 & $\{11, 22, 5, 2, 2\}$ & 56 & 28 & 25\\
	23 & $\{3, 11, 18, 3, 2\}$ & -132 & 66 & 10 & 116 & $\{12, 3, 5, 165, 2\}$ & 120 & 60 & 0\\
	24 & $\{3, 14, 4, 13, 2\}$ & 240 & 120 & 6 & 117 & $\{12, 3, 8, 9, 2\}$ & -288 & 144 & 5\\
	25 & $\{3, 15, 5, 6, 2\}$ & 36 & 18 & 48 & 118 & $\{12, 9, 3, 7, 2\}$ & -336 & 168 & 0\\
	26 & $\{3, 18, 11, 3, 2\}$ & 132 & 66 & 10  & 119 & $\{12, 148, 4, 2, 2\}$ & 24 & 12 & 0\\
	27 & $\{3, 25, 5, 5, 2\}$ & 90 & 45 & 37 & 120 & $\{12, 149, 3, 3, 2\}$ & 48 & 24 & 0\\
	28 & $\{3, 29, 9, 3, 2\}$ & 90 & 45 & 23 & 121 & $\{13, 3, 5, 89, 2\}$ & 234 & 117 & 5\\
	29 & $\{3, 51, 8, 3, 2\}$ & 48 & 24 & 9 & 122 & $\{13, 3, 17, 5, 2\}$ & -330 & 165 & 5\\
	30 & $\{3, 53, 4, 7, 2\}$ & 72 & 36 & 9 & 123 & $\{13, 4, 14, 3, 2\}$ & -240 & 120 & 6\\
	31 & $\{4, 3, 9, 117, 2\}$ & 72 & 36 & 3 & 124 & $\{13, 5, 155, 2, 2\}$ & 20 & 10 & 0\\
	32 & $\{4, 3, 11, 29, 2\}$ & 216 & 108 & 13 & 125 & $\{13, 6, 3, 18, 2\}$ & 312 & 156 & 6\\
	33 & $\{4, 3, 14, 17, 2\}$ & -168 & 84 & 11 & 126 & $\{13, 81, 3, 3, 2\}$ & 138 & 69 & 9\\
	34 & $\{4, 3, 36, 9, 2\}$ & -24 & 12 & 21 & 127 & $\{14, 3, 16, 5, 2\}$ & 96 & 48 & 51\\
	35 & $\{4, 4, 6, 20, 2\}$ & 144 & 72 & 25 & 128 & $\{14, 5, 3, 163, 2\}$ & 84 & 42 & 0\\
	36 & $\{4, 4, 86, 4, 2\}$ & -8 & 4 & 4 & 129 & $\{14, 5, 7, 3, 2\}$ & -156 & 78 & 21\\
	37 & $\{4, 6, 6, 6, 2\}$ & 240 & 120 & 0 & 130 & $\{14, 6, 16, 2, 2\}$ & 264 & 132 & 4\\
	38 & $\{4, 7, 6, 5, 2\}$ & 0 & 0 & 55 & 131 & $\{15, 3, 5, 51, 2\}$ & 330 & 165 & 11\\
	39 & $\{4, 10, 4, 8, 2\}$ & -184 & 92 & 12 & 132 & $\{15, 3, 9, 7, 2\}$ & 54 & 27 & 35\\
	40 & $\{4, 14, 146, 2, 2\}$ & 56 & 28 & 0 & 133 & $\{15, 5, 3, 87, 2\}$ & 210 & 105 & 11\\
	41 & $\{4, 15, 3, 153, 2\}$ & 24 & 12 & 0 & 134 & $\{15, 47, 3, 3, 2\}$ & 210 & 105 & 7\\
	42 & $\{4, 18, 38, 2, 2\}$ & -144 & 72 & 7 & 135 & $\{16, 6, 14, 2, 2\}$ & -264 & 132 & 4\\
	43 & $\{4, 21, 6, 3, 2\}$ & 96 & 48 & 30 & 136 & $\{16, 40, 4, 2, 2\}$ & -160 & 80 & 18\\
	44 & $\{4, 25, 3, 23, 2\}$ & -336 & 168 & 0 & 137 & $\{17, 5, 3, 49, 2\}$ & 306 & 153 & 5\\
	45 & $\{4, 27, 3, 21, 2\}$ & 408 & 204 & 0 & 138 & $\{17, 14, 3, 4, 2\}$ & 168 & 84 & 11\\
	46 & $\{4, 40, 16, 2, 2\}$ & 160 & 80 & 18 & 139 & $\{18, 3, 6, 13, 2\}$ & -312 & 156 & 6\\
	47 & $\{4, 85, 5, 3, 2\}$ & 0 & 0 & 3 & 140 & $\{19, 3, 5, 32, 2\}$ & 456 & 228 & 0\\
	48 & $\{4, 86, 4, 4, 2\}$ & 8 & 4 & 4 & 141 & $\{19, 4, 11, 3, 2\}$ & 144 & 72 & 11\\
	49 & $\{4, 148, 12, 2, 2\}$ & -24 & 12 & 0 & 142 & $\{19, 30, 3, 3, 2\}$ & 300 & 150 & 0\\
	50 & $\{4, 157, 3, 11, 2\}$ & 48 & 24 & 0 & 143 & $\{20, 3, 13, 5, 2\}$ & -240 & 120 & 14\\
	51 & $\{5, 3, 5, 6, 3\}$ & 186 & 93 & 6 & 144 & $\{20, 4, 4, 10, 2\}$ & 136 & 68 & 19\\
	52 & $\{5, 3, 34, 7, 2\}$ & -156 & 78 & 14 & 145 & $\{20, 6, 4, 4, 2\}$ & -144 & 72 & 25\\
	53 & $\{5, 3, 115, 6, 2\}$ & -36 & 18 & 2 & 146 & $\{21, 3, 27, 4, 2\}$ & -408 & 204 & 0\\
	54 & $\{5, 4, 6, 11, 2\}$ & -120 & 60 & 25 & 147 & $\{21, 5, 3, 30, 2\}$ & 420 & 210 & 0\\
	55 & $\{5, 5, 7, 5, 2\}$ & 86 & 43 & 34 & 148 & $\{21, 8, 3, 6, 2\}$ & -240 & 120 & 19\\
	56 & $\{5, 5, 25, 3, 2\}$ & -90 & 45 & 37 & 149 & $\{22, 3, 3, 6, 3\}$ & 84 & 42 & 19\\
	57 & $\{5, 6, 7, 4, 2\}$ & 0 & 0 & 55 & 150 & $\{22, 3, 6, 3, 3\}$ & -84 & 42 & 19\\
	58 & $\{5, 7, 5, 5, 2\}$ & -86 & 43 & 34 & 151 & $\{23, 3, 25, 4, 2\}$ & 336 & 168 & 0\\
	59 & $\{5, 8, 3, 4, 3\}$ & -120 & 60 & 21 & 152 & $\{23, 5, 25, 2, 2\}$ & 340 & 170 & 0\\
	60 & $\{5, 9, 7, 3, 2\}$ & 30 & 15 & 45 & 153 & $\{24, 5, 6, 3, 2\}$ & -120 & 60 & 34 \\
	61 & $\{5, 9, 115, 2, 2\}$ & 36 & 18 & 2 & 154 & $\{25, 5, 23, 2, 2\}$ & -340 & 170 & 0\\
	62 & $\{5, 10, 3, 62, 2\}$ & 120 & 60 & 31 & 155 & $\{28, 21, 3, 3, 2\}$ & -336 & 168 & 0\\
	63 & $\{5, 11, 3, 34, 2\}$ & 180 & 90 & 22 & 156 & $\{29, 11, 3, 4, 2\}$ & -216 & 108 & 13\\
	64 & $\{5, 13, 3, 20, 2\}$ & 240 & 120 & 14 & 157 & $\{30, 3, 5, 21, 2\}$ & -420 & 210 & 0\\
	65 & $\{5, 16, 3, 14, 2\}$ & -96 & 48 & 51 & 158 & $\{31, 6, 3, 9, 2\}$ & -252 & 126 & 17\\
	66 & $\{5, 17, 3, 13, 2\}$ & 330 & 165 & 5 & 159 & $\{32, 5, 3, 19, 2\}$ & -456 & 228 & 0\\
	67 & $\{5, 22, 11, 2, 2\}$ & -56 & 28 & 25 & 160 & $\{34, 3, 11, 5, 2\}$ & -180 & 90 & 22\\
	68 & $\{5, 23, 3, 10, 2\}$ & -36 & 18 & 42 & 161 & $\{38, 18, 4, 2, 2\}$ & 144 & 72 & 7\\
	69 & $\{5, 37, 3, 8, 2\}$ & 24 & 12 & 34 & 162 & $\{45, 17, 3, 3, 2\}$ & -246 & 123 & 9\\
	70 & $\{5, 65, 3, 7, 2\}$ & 114 & 57 & 25 & 163 & $\{47, 7, 4, 3, 2\}$ & -96 & 48 & 14\\
	71 & $\{6, 3, 4, 7, 3\}$ & 72 & 36 & 33 & 164 & $\{49, 3, 5, 17, 2\}$ & -306 & 153 & 5\\
	72 & $\{6, 3, 8, 21, 2\}$ & 240 & 120 & 19 & 165 & $\{51, 5, 3, 15, 2\}$ & -330 & 165 & 11\\
	73 & $\{6, 5, 3, 5, 3\}$ & -186 & 93 & 6 & 166 & $\{59, 3, 7, 7, 2\}$ & -126 & 63 & 21\\
	74 & $\{6, 5, 15, 3, 2\}$ & -36 & 18 & 48 & 167 & $\{62, 3, 10, 5, 2\}$ & -120 & 60 & 31 \\
	75 & $\{6, 6, 6, 4, 2\}$ & -240 & 120 & 0 & 168 & $\{64, 3, 6, 9, 2\}$ & -144 & 72 & 17 \\
	76 & $\{6, 13, 3, 11, 2\}$ & -252 & 126 & 8 & 169 & $\{74, 3, 3, 5, 3\}$ & -6 & 3 & 6 \\
	77 & $\{6, 22, 8, 2, 2\}$ & 144 & 72 & 28 & 170 & $\{74, 3, 5, 3, 3\}$ & 6 & 3 & 6\\
	78 & $\{6, 112, 6, 2, 2\}$ & 0 & 0 & 1 & 171 & $\{79, 15, 3, 3, 2\}$ & -174 & 87 & 7 \\
	79 & $\{6, 115, 3, 5, 2\}$ & 36 & 18 & 2 & 172 & $\{87, 3, 5, 15, 2\}$ & -210 & 105 & 11\\
	80 & $\{7, 3, 6, 123, 2\}$ & 84 & 42 & 3 & 173 & $\{89, 5, 3, 13, 2\}$ & -234 & 117 & 5\\
	81 & $\{7, 3, 9, 12, 2\}$ & 336 & 168 & 0 & 174 & $\{97, 4, 8, 3, 2\}$ & -24 & 12 & 5\\
	82 & $\{7, 3, 65, 5, 2\}$ & -114 & 57 & 25 & 175 & $\{101, 4, 4, 7, 2\}$ & -56 & 28 & 4\\
	83 & $\{7, 4, 3, 6, 3\}$ & -72 & 36 & 33 & 176 & $\{115, 9, 5, 2, 2\}$ & -36 & 18 & 2\\
	84 & $\{7, 4, 4, 101, 2\}$ & 56 & 28 & 4 & 177 & $\{117, 9, 3, 4, 2\}$ & -72 & 36 & 3\\
	85 & $\{7, 4, 53, 3, 2\}$ & -72 & 36 & 9 & 178 & $\{118, 6, 8, 2, 2\}$ & -24 & 12 & 2\\
	86 & $\{7, 7, 3, 59, 2\}$ & 126 & 63 & 21 & 179 & $\{123, 6, 3, 7, 2\}$ & -84 & 42 & 3\\
	87 & $\{7, 9, 3, 15, 2\}$ & -54 & 27 & 35 & 180 & $\{146, 14, 4, 2, 2\}$ & -56 & 28 & 0\\
	88 & $\{7, 10, 5, 3, 2\}$ & 180 & 90 & 6 & 181 & $\{147, 14, 3, 3, 2\}$ & -84 & 42 & 0\\
	89 & $\{7, 12, 3, 9, 2\}$ & 336 & 168 & 0 & 182 & $\{153, 3, 15, 4, 2\}$ & -24 & 12 & 0\\
	90 & $\{7, 13, 9, 2, 2\}$ & 84 & 42 & 39 & 183 & $\{155, 5, 13, 2, 2\}$ & -20 & 10 & 0\\
	91 & $\{7, 34, 3, 5, 2\}$ & 156 & 78 & 14 & 184 & $\{163, 3, 5, 14, 2\}$ & -84 & 42 & 0\\
	92 & $\{8, 3, 10, 9, 2\}$ & 120 & 60 & 35 & 185 & $\{165, 5, 3, 12, 2\}$ & -120 & 60 & 0\\
	93 & $\{8, 3, 37, 5, 2\}$ & -24 & 12 & 34 & & & & & \\
	\hline 
	
\end{longtable}

\newpage
\section*{Appendix B}

\begin{table}[!h]
	
	\caption{\normalsize Number of generations and singlets for the case $J_A \subset SL_W$. 
		Here, $\text{diag}(A)$ denotes the diagonal elements of the loop polynomial exponent matrix \eqref{eq:matrixA}. 
		In this case, the subgroups $G_1$ and $G_2$ (and their respective mirror dual pairs, $G_1^T$ and $G_2^T$) are obtained from the maximal admissible group $SL_W$. 
		The subgroups $G_1$ and $G_2^T$ are $\mathbb{Z}_1$; however, the corresponding subclasses of invariant deformations differ (a similar situation applies to the subgroups $G_2$ and $G_1^T$). 
		The Euler characteristic $\chi$ for each case is calculated using Vafa's formula \eqref{eq:Vafa}. Note that the number of generations $N_{\text{gen}}$, computed as the difference in the number of deformations \eqref{eq:deform}, is related to the Euler characteristic via $|\chi| = 2N_{\text{gen}}$, indicating the absence of contributions from Roan pairs. The number of singlets is obtained using formula \eqref{eq:singlets}.}
	\label{tab2}
	\centering
	\footnotesize
	
	\begin{tabular}{|c|c|c|c|c|c|c|}
		\hline
		\textbf{} & \textbf{} & \textbf{Groups} & \textbf{Groups} & \textbf{} & \textbf{$N_{\text{gen}}$} & \textbf{Singlets} \\
		\textbf{№} & \textbf{$\text{diag}(A)$} & \textbf{$(G_1,G^T_2)$} & \textbf{$(G_1,G^T_2)$} & $\left(\chi(G_1),\chi{(G_2)}\right)$&\textbf{$(G_1,G^T_2)$} & \textbf{$(G_1,G^T_2)$} \\
		\hline
	1 & $\{4, 4, 4, 4, 4\}$ & $\mathbb{Z}_1$ & $\mathbb{Z}_{41}$ & $(-200, 200)$ & $(100, 100)$ & $(0, 0)$ \\
	2 & $\{4, 16, 56, 2, 2\}$ & $\mathbb{Z}_1$ & $\mathbb{Z}_3$ & $(-96, 184)$ & $(48, 92)$ & $(18, 11)$ \\
	3 & $\{4, 22, 26, 2, 2\}$ & $\mathbb{Z}_1$ & $\mathbb{Z}_3$ & $(32, 312)$ & $(16, 156)$ & $(29, 0)$ \\
	4 & $\{4, 28, 20, 2, 2\}$ & $\mathbb{Z}_1$ & $\mathbb{Z}_3$ & $(-280, -16)$ & $(140, 8)$ & $(0, 18)$ \\
	5 & $\{4, 58, 14, 2, 2\}$ & $\mathbb{Z}_1$ & $\mathbb{Z}_3$ & $(-152, 112)$ & $(76, 56)$ & $(11, 7)$ \\
	6 & $\{5, 4, 7, 8, 2\}$ & $\mathbb{Z}_1$ & $\mathbb{Z}_3$ & $(56, 240)$ & $(28, 120)$ & $(44, 3)$ \\
	7 & $\{5, 5, 4, 31, 2\}$ & $\mathbb{Z}_1$ & $\mathbb{Z}_3$ & $(-64, 120)$ & $(32, 60)$ & $(22, 27)$ \\
	8 & $\{5, 8, 4, 7, 2\}$ & $\mathbb{Z}_1$ & $\mathbb{Z}_3$ & $(56, 240)$ & $(28, 120)$ & $(36, 3)$ \\
	9 & $\{5, 10, 43, 2, 2\}$ & $\mathbb{Z}_1$ & $\mathbb{Z}_3$ & $(-80, 136)$ & $(40, 68)$ & $(24, 17)$ \\
	10 & $\{5, 13, 19, 2, 2\}$ & $\mathbb{Z}_1$ & $\mathbb{Z}_3$ & $(-156, 20)$ & $(78, 10)$ & $(10, 39)$ \\
	11 & $\{7, 4, 5, 11, 2\}$ & $\mathbb{Z}_1$ & $\mathbb{Z}_3$ & $(40, 224)$ & $(20, 112)$ & $(34, 9)$ \\
	12 & $\{7, 4, 8, 5, 2\}$ & $\mathbb{Z}_1$ & $\mathbb{Z}_3$ & $(-240, -56)$ & $(120, 28)$ & $(3, 36)$ \\
	13 & $\{7, 7, 41, 2, 2\}$ & $\mathbb{Z}_1$ & $\mathbb{Z}_3$ & $(-84, 124)$ & $(42, 62)$ & $(33, 26)$ \\
	14 & $\{8, 4, 4, 38, 2\}$ & $\mathbb{Z}_1$ & $\mathbb{Z}_3$ & $(-72, 160)$ & $(36, 80)$ & $(18, 18)$ \\
	15 & $\{8, 7, 4, 5, 2\}$ & $\mathbb{Z}_1$ & $\mathbb{Z}_3$ & $(-240, -56)$ & $(120, 28)$ & $(3, 44)$ \\
	16 & $\{8, 10, 10, 2, 2\}$ & $\mathbb{Z}_1$ & $\mathbb{Z}_3$ & $(64, 280)$ & $(32, 140)$ & $(42, 0)$ \\
	17 & $\{10, 10, 8, 2, 2\}$ & $\mathbb{Z}_1$ & $\mathbb{Z}_3$ & $(-280, -64)$ & $(140, 32)$ & $(0, 42)$ \\
	18 & $\{11, 4, 4, 17, 2\}$ & $\mathbb{Z}_1$ & $\mathbb{Z}_3$ & $(32, 264)$ & $(16, 132)$ & $(31, 8)$ \\
	19 & $\{11, 5, 4, 7, 2\}$ & $\mathbb{Z}_1$ & $\mathbb{Z}_3$ & $(-224, -40)$ & $(112, 20)$ & $(9, 34)$ \\
	20 & $\{11, 7, 13, 2, 2\}$ & $\mathbb{Z}_1$ & $\mathbb{Z}_3$ & $(44, 252)$ & $(22, 126)$ & $(53, 6)$ \\
	21 & $\{13, 7, 11, 2, 2\}$ & $\mathbb{Z}_1$ & $\mathbb{Z}_3$ & $(-252, -44)$ & $(126, 22)$ & $(6, 53)$ \\
	22 & $\{14, 58, 4, 2, 2\}$ & $\mathbb{Z}_1$ & $\mathbb{Z}_3$ & $(-112, 152)$ & $(56, 76)$ & $(7, 11)$ \\
	23 & $\{17, 4, 4, 11, 2\}$ & $\mathbb{Z}_1$ & $\mathbb{Z}_3$ & $(-264, -32)$ & $(132, 16)$ & $(8, 31)$ \\
	24 & $\{19, 13, 5, 2, 2\}$ & $\mathbb{Z}_1$ & $\mathbb{Z}_3$ & $(-20, 156)$ & $(10, 78)$ & $(39, 10)$ \\
	25 & $\{20, 28, 4, 2, 2\}$ & $\mathbb{Z}_1$ & $\mathbb{Z}_3$ & $(16, 280)$ & $(8, 140)$ & $(18, 0)$ \\
	26 & $\{26, 22, 4, 2, 2\}$ & $\mathbb{Z}_1$ & $\mathbb{Z}_3$ & $(-312, -32)$ & $(156, 16)$ & $(0, 29)$ \\
	27 & $\{31, 4, 5, 5, 2\}$ & $\mathbb{Z}_1$ & $\mathbb{Z}_3$ & $(-120, 64)$ & $(60, 32)$ & $(27, 22)$ \\
	28 & $\{38, 4, 4, 8, 2\}$ & $\mathbb{Z}_1$ & $\mathbb{Z}_3$ & $(-160, 72)$ & $(80, 36)$ & $(18, 18)$ \\
	29 & $\{41, 7, 7, 2, 2\}$ & $\mathbb{Z}_1$ & $\mathbb{Z}_3$ & $(-124, 84)$ & $(62, 42)$ & $(26, 33)$ \\
	30 & $\{43, 10, 5, 2, 2\}$ & $\mathbb{Z}_1$ & $\mathbb{Z}_3$ & $(-136, 80)$ & $(68, 40)$ & $(17, 24)$ \\
	31 & $\{56, 16, 4, 2, 2\}$ & $\mathbb{Z}_1$ & $\mathbb{Z}_3$ & $(-184, 96)$ & $(92, 48)$ & $(11, 18)$ \\
	\hline
	\end{tabular}
	
\end{table}

\end{document}